\documentclass[11pt,twoside]{article}
\usepackage{asp2006}
\usepackage{psfig}
\usepackage{epsf}
\usepackage{graphics}
\usepackage{graphicx}
\usepackage{lscape}
\begin{document}

\title{Star-powered LINERs in the Sloan Digital Sky Survey}
\author{N. Vale Asari$^{1,2}$,
  G. Stasi\'nska$^{1}$, 
  R. Cid Fernandes$^{2,1}$,
  J. M. Gomes$^{3,1}$, \\
  M. Schlickmann$^{2}$,
  A. Mateus$^{4}$, 
  W. Schoenell$^{2}$,
  L. Sodr\'e Jr.$^{4}$ \\
  (the SEAGal collaboration)
}
\affil{\footnotesize
  $^{1}$LUTH, Observatoire de Paris, CNRS, Universit\'e Paris Diderot;
  Place Jules Janssen 92190 Meudon, France\\
  $^{2}$Departamento de F\'{\i}sica, CFM, UFSC, Florian\'opolis, SC, Brazil\\
  $^{3}$GEPI, Observatoire de Paris, CNRS, Universit\'e Paris Diderot;
  Place Jules Janssen 92190 Meudon, France\\
  $^{4}$Instituto de Astronomia, Geof\'{\i}sica e Ci\^encias
  Atmosf\'ericas, USP, SP,  Brazil
}

\begin{abstract}
  Galaxies are usually classified as star forming or active by using
  diagnostic diagrams, such as [N~{\sc ii}]/H$\alpha$ vs.\ [O~{\sc
    iii}]/H$\beta$.  Active galaxies are further classified into
  Seyfert or LINER-like sources. We claim that a non-negligible
  fraction of galaxies classified as LINERs in the Sloan Digital Sky
  Survey are in fact ionized by hot post-AGB stars and white
  dwarfs.
\end{abstract}

\section{Background, sample and data processing}

Heckman (1980) was the first to classify as Low Ionization Nuclear
Emission-Line Regions (LINERs) those active nuclei whose optical
emission lines have similar widths to Seyfert galaxies but show lower
excitation.  There has never been a consensus whether every LINER has
an active galactic nucleus (AGN).

\begin{figure}[!b]
  \begin{center}
    \includegraphics[bb=40 490 592 670,width=0.65\textwidth]{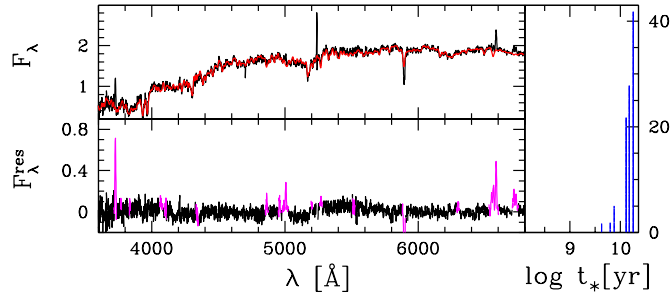}
  \end{center}
  \caption{Example of {\sc starlight} spectral fit to a SDSS galaxy
    with LINER-like emission line ratios.}
\label{fig:fits}
\end{figure}

The Sloan Digital Sky Survey (SDSS; York et al. 2000) has provided the
astronomical community with thousands of LINER-like spectra.  Kewley
et al. (2006) showed that, in the [N~{\sc ii}]/H$\alpha$ vs.\ [O~{\sc
    iii}]/H$\beta$ plane (BPT, after Baldwin, Philips \& Terlevich
1981), the right ``wing'' of SDSS galaxies is in fact composed of two
branches, which they identified as Seyferts and LINERs. The fraction
of objects classified as LINERs in the local Universe by diagnostic
diagram methods depends on the dividing line adopted to separate
star-forming (SF) galaxies (the left ``wing'') from the galaxies with
an AGN (the right ``wing'').

Trinchieri \& di Serego Alighieri (1991) and Binette et al.\ (1994)
suggested that hot post-AGB stars and white dwarfs can explain the
emission lines observed in elliptical galaxies with little gas and no
current star formation.  Here we explore this idea in more detail by
using the stellar population results obtained directly by our
synthesis analysis for SDSS galaxies and check whether galaxies that
have stopped forming stars --- dubbed {\em retired} galaxies --- can
explain the emission line spectra of LINER-like galaxies.

Our sample consists of 131287 SDSS galaxies. We infer the stellar
populations that compose a galaxy by a pixel-by-pixel model of its
spectral continuum with our code {\sc starlight}, using a base of
simple stellar populations from Bruzual \& Charlot (2003). The top
panel of Fig.~\ref{fig:fits} shows an example of observed (black) and
model (red) spectra.  The panel on the right shows the light fraction
associated with populations of different ages for this model.  We
measure emission lines by fitting gaussians to the residual spectrum
(bottom panel).

\begin{figure}[!t]
  \begin{center}
    \includegraphics[bb= 14 55 563 310,width=0.7\textwidth]{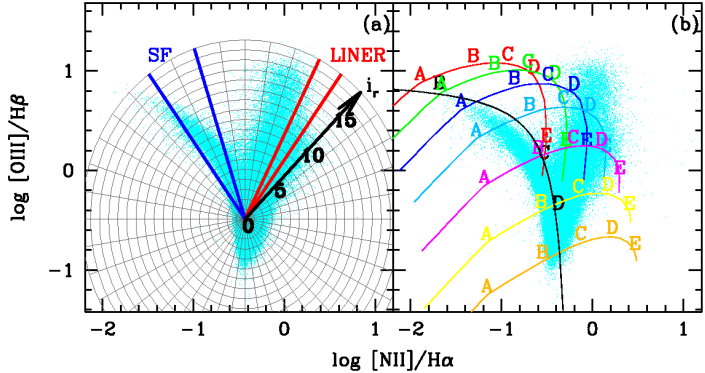}
  \end{center}
  \caption{(a) Galaxy sample on the BPT diagram, chopped in polar
    coordinates. (b) Photoionization model sequences using the
    radiation inferred by {\sc starlight} for the LINER branch, shown
    for different values of $\log U$: -2.3 (red), -2.7 (green), -3
    (blue), -3.3 (cyan), -3.7 (purple), -4 (yellow), -4.4
    (orange). The black line is the dividing line for AGN and SF
    galaxies according to Stasi\'nska et al. (2006). The metallicities
    $Z/Z_\odot$ are marked with letters as follows: 0.2 (A), 0.5 (B),
    1 (C), 2 (D), 5 (E).}
\label{fig:BPT}
\end{figure}

In Fig.~\ref{fig:BPT}a, we show our galaxy sample on the BPT plane.
We chop the BPT in bins defined by their polar coordinates ($r,
\theta$), with the center at the point of inflection of the median
curve of the distribution of [O~{\sc iii}]/H$\beta$ as a function of
[N~{\sc ii}]/H$\alpha$.  We study variations of properties in the $r$
coordinate by using the $i_r$ index for the SF and LINER branches as
defined in Fig.~\ref{fig:BPT}a.

\section{Models and discussion}

\begin{figure}[!ht]
  \begin{center}
    \includegraphics[bb=40 200 580 690, width=0.6\textwidth]{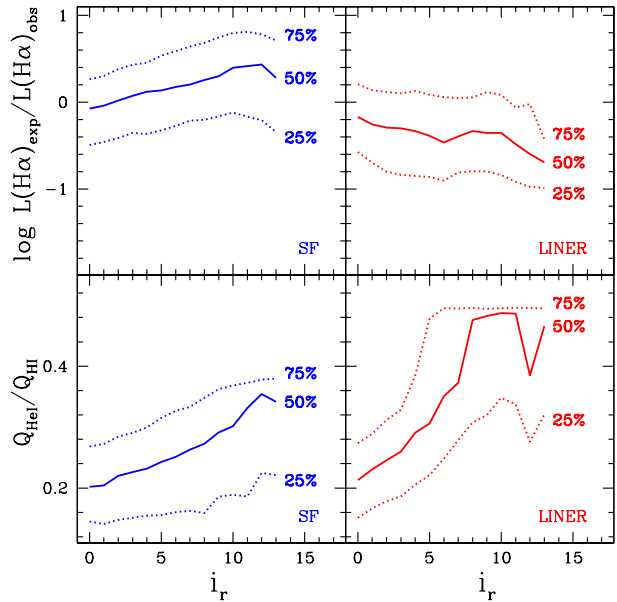}
  \end{center}
  \caption{Variation with $i_r$ of the ratio of the expected to
    observed $L({\rm H}\alpha)$ (top) and of the hardness ratio
    (bottom) for the SF and LINER branches.}
\label{fig:LHa}
\end{figure}

For each galaxy, we compute $Q_{\rm HI}$, the number of stellar
ionizing photons with energy above 13.6 eV arising from the
populations uncovered by {\sc starlight}. We predict the luminosity in
H$\alpha$ due to the stellar populations, $L({\rm H}\alpha)_{\rm
  exp}$, assuming that all these photons are absorbed by the gas in
the galaxy, and compare it to the observed value $L({\rm
  H}\alpha)_{\rm obs}$ (top panels of Fig.~\ref{fig:LHa}).  The
$L({\rm H}\alpha)_{\rm obs}$ for the SF branch is well explained by
the populations recovered by {\sc starlight}.  We can explain the
observed $L({\rm H}\alpha)$ for about 25\% of the LINER branch in our
sample by using the ionization of old stellar populations only
(however, see Stasi{\'n}ska et al.\ 2008 for a more detailed
discussion). This fraction increases substantially if one allows
objects with lower $S/N$ emission lines into the sample.

Bottom panels of Fig.~\ref{fig:LHa} show the ratio $Q_{\rm
  HeI}/Q_{\rm HI}$ of He to H ionizing photons (expressing the
hardness of the radiation field) for the SF and LINER branches.  The
stars in the LINER branch have much harder radiation than the ones in
the SF branch, so that we expect the [O~{\sc iii}] and [N~{\sc ii}]
lines to be stronger with respect to hydrogen lines in LINERs than in
SF galaxies.

Photoionization models confirm that the ionization from old stars can
account for the emission line ratios in the LINER branch.
Fig.~\ref{fig:BPT}b shows models for given values of
nebular metallicity and ionization parameter. Metal-rich models (twice
the solar metallicity) are the ones which best cover the LINER region.

A full version of this study is presented in Stasi{\'n}ska et
al.\ (2008).

\end{document}